\documentclass[12pt]{article}
  \usepackage{amsfonts}
  \usepackage{amsmath}
\usepackage{amssymb}
\usepackage{amscd}
\usepackage[dvips]{graphicx}
\begin{document}
~~
\bigskip
\bigskip
\begin{center}
{\Large {\bf{{{Oscillator model on  Lie-algebraically deformed
nonrelativistic  space-time}}}}}
\end{center}
\bigskip
\bigskip
\bigskip
\begin{center}
{{\large ${\rm {Marcin\;Daszkiewicz^{1}}}$, \large ${\rm {Cezary\;
 J.\;Walczyk^{2}}}$}}
\end{center}
\bigskip
\begin{center}
\bigskip

{ ${\rm{~^{1}Institute\; of\; Theoretical\; Physics}}$}

{ ${\rm{ University\; of\; Wroclaw\; pl.\; Maxa\; Borna\; 9,\;
50-206\; Wroclaw,\; Poland}}$}

{ ${\rm{ e-mail:\; marcin@ift.uni.wroc.pl}}$}

\bigskip

{ ${\rm{~^{2}Department\; of\; Physics}}$}

{ ${\rm{ University\; of\; Bialystok,\; ul.\; Lipowa\; 41,\;
15-424\;Bialystok,\; Poland}}$}

{ ${\rm{ e-mail:\; c.walczyk@alpha.uwb.edu.pl}}$}

\end{center}
\bigskip
\bigskip
\bigskip
\bigskip
\bigskip
\bigskip
\bigskip
\bigskip
\bigskip
\begin{abstract}
The classical and quantum oscillator model on  Lie-algebraically
deformed nonrelativistic space-time is introduced and analyzed. The
corresponding equations of motions are studied using mostly
numerical methods. The time-dependent energy spectrum is presented
as well.
\end{abstract}
\bigskip
\bigskip
\bigskip
\bigskip
\eject

\section{{{Introduction}}}

~~~\,\,\,Due to several  theoretical arguments (see e.g.
\cite{grav1}-\cite{string2}) the interest in  studying of space-time
noncommutativity is growing rapidly. There appeared a lot of papers
dealing with noncommutative classical (\cite{mech1}-\cite{daszwal})
and quantum (\cite{qm0}-\cite{giri}) mechanics, as well as with
field theoretical models (see e.g. (\cite{fieldus}-\cite{fieldn}),
defined on quantum
 space-time.

At present, in accordance with Hopf-algebraic classification  of all
deformations of relativistic and nonrelativistic symmetries (see
\cite{zakrzewski}, \cite{kowclas}) one can distinguish two quite
interesting  kinds of  quantum spaces. First of them corresponds to
the well-known canonical type of noncommutativity

\begin{equation}
[\;{\hat x}_{\mu},{\hat x}_{\nu}\;] =
i\theta_{\mu\nu}\;,
\label{wielkaslawia}
\end{equation}
\\
with antisymmetric constant tensor $\theta^{\mu\nu}$. Its
relativistic and nonrelativistic Hopf-algebraic counterparts have
been proposed in \cite{3e}-\cite{3b} and \cite{daszgali} respectively\footnote{See also \cite{daszdual}.}.\\
The second class of mentioned deformations  introduces the
Lie-algebraic type of space-time noncommutativity

\begin{equation}
[\;{\hat x}_{\mu},{\hat x}_{\nu}\;] = i\theta_{\mu\nu}^{\rho}{\hat
x}_{\rho}\;, \label{noncomm1}
\end{equation}
\\
with particularly chosen constant coefficients
$\theta_{\mu\nu}^{\rho}$. The corresponding Poincare quantum groups
have been introduced in \cite{4a}, \cite{lukwor}, while the suitable
Galilei algebras  - in \cite{kappaga} and \cite{daszgali}.

 Recently, there was
proposed  a particular type of  Lie-algebraic deformation of
nonrelativistic space-time with spatial directions commuting to
time
\footnote{It will be call   $t$-deformation of classical space.}

\begin{equation}
[\;t,{ {\hat x}}_{i}\;] = 0 = [\;{ {\hat x}}_{k},{{\hat
x}}_{\rho(\tau)}\;]\;\;\;,\;\;\;[\;{ {\hat x}}_{\tau},{ {\hat
x}}_{\rho}\;] = \frac{i}{\kappa}t\;\;\;\;;\;\;\;\rho, \tau, k - {\rm
fixed\;\;and\;\;different}\;.
 \label{wstep1}
\end{equation}
\\
The above noncommutativity has been obtained in the framework of
quantum groups in \cite{daszgali}, 
while its basic  properties 
have been investigated in a context of nonrelativistic particle
subjected to the external constant force  \cite{daszwal}. In
particular, there was demonstrated that such a kind of quantum
space-time produces additional acceleration of moving  particle
 coupled to the force terms generated by so-called space-like
deformation  \cite{daszgali}.

In this article we extend our investigations of deformation
(\ref{wstep1}) to  more complicated nonrelativistic system - the
classical and quantum oscillator model. It should be noted, however,
that analogous studies have been already performed in a context of
canonical deformation  (\ref{wielkaslawia}), i.e. the solution of
corresponding  equation of motion has been provided and analyzed in
\cite{romeroads}, while the (deformed) energy spectrum has been
discussed in  \cite{lumom}-
\cite{jellal}. 

In this paper we adopt the general treatment proposed in
\cite{kijanka} and show, that the space-time noncommutativity
(\ref{wstep1}) generates a proper (explicit) time-dependence of
oscillator Hamiltonian function.
 In such a  way we discover a
 connection between noncommutative geometry (a quantum group)
and nonrelativistic models with time-dependent mass and frequency
(see e.g. \cite{slawia}-\cite{slawia10}). Besides, we also confirm
that such a deformation introduces the additional term proportional
to angular momentum of considered system  \cite{lumom},
\cite{kijanka}. Finally, the numerical studies of the
 solutions of corresponding  equations of motion are
presented in detail and the growing in time  energy spectrum  is
analyzed as well\footnote{It should be noted that for fixed time
parameter $t$ the obtained  energy spectrum becomes the same as one
recovered in \cite{kijanka}.}.

In this article apart of deformation (\ref{wstep1}) we also
introduced its natural generalization

\begin{equation}
[\;t,{ {\hat x}}_{i}\;] = 0 = [\;{ {\hat x}}_{k},{{\hat
x}}_{\rho(\tau)}\;]\;\;\;,\;\;\;[\;{ {\hat x}}_{\tau},{ {\hat
x}}_{\rho}\;] = if_{\kappa}(t)\;, \label{wstep2}
\end{equation}
\\
with arbitrary time-dependent function $f_{\kappa}(t)$ approaching
zero  for  parameter $\kappa$ running to infinity. We demonstrate
that for such generalized space-time the energy spectrum of the
model depends on function $f_{\kappa}(t)$ and becomes finite for
large times. It should be noted however, that noncommutativity
(\ref{wstep2}) can not be derived in the framework of quantum
groups, and for this reason, from formal point of view its geometric
status remains unknown.

The paper is organized as follows. In second Section we recall the
Galilei Hopf structure providing quantum space (\ref{wstep1}).
Further, in  Section 3, we investigate the classical oscillator
model on such a space-time, i.e. the corresponding equation of
motion is provided and its solution is analyzed numerically as well.
In Section 4 the energy spectrum of a proper quantum model is
discovered. Section 5 deals with  the classical and quantum
oscillator system defined on generalized space-time (\ref{wstep2}).
The results are summarized and discussed in the last Section.

\section{{{Twisted Galilei Hopf algebra and corresponding Lie-algebraically deformed space-time}}}

In this Section (following the paper \cite{daszgali}) we recall the
Lie-algebraically deformed Galilei Hopf algebra ${\mathcal
U}_{\kappa}(\mathcal{G})$ ($\kappa$ denotes deformation parameter)
providing nonrelativistic space-time (\ref{wstep1}). Its algebraic
sector remains classical\footnote{The symbols $K_{ij}$, $V_i$ and
$\Pi_\mu$ denote rotations, boosts  and space-time translation
generators respectively.}
\begin{eqnarray}
&&\left[\, K_{ij},K_{kl}\,\right] =i\left( \delta
_{il}\,K_{jk}-\delta
_{jl}\,K_{ik}+\delta _{jk}K_{il}-\delta _{ik}K_{jl}\right) \;,  \notag \\
&~~&  \cr &&\left[\, K_{ij},V_{k}\,\right] =i\left( \delta
_{jk}\,V_i-\delta _{ik}\,V_j\right)\;\; \;, \;\;\;\left[
\,K_{ij},\Pi_{k }\,\right] =i\left( \delta _{j k }\,\Pi_{i }-\delta
_{ik }\,\Pi_{j }\right) \;, \label{nnnga}
\\
&~~&  \cr &&\left[ \,K_{ij},\Pi_{0 }\,\right] =\left[
\,V_i,V_j\,\right] = \left[ \,V_i,\Pi_{j }\,\right]
=0\;\;\;,\;\;\;\left[ \,V_i,\Pi_{0 }\,\right]
=-i\Pi_i\;\;\;,\;\;\;\left[ \,\Pi_{\mu },\Pi_{\nu }\,\right] =
0\;,\nonumber
\end{eqnarray}
\\
while the coalgebraic part takes the form ($\rho$, $\tau$ - fixed
and  different)

\begin{eqnarray}
 \Delta_{{\kappa}}(\Pi_0)&=&\Delta _0(\Pi_0) +
\frac{1}{2{{\kappa}}} \Pi_\tau \wedge \Pi_\rho\;,\label{gacoa1}\\
 &~~&  \cr
\Delta_{{\kappa}}(\Pi_i)&=&\Delta
_0(\Pi_i)\;\;\;,\;\;\;\Delta_{{\kappa}}(V_i)=\Delta_0(V_i)\;,\label{coa0}\\
 &~~&  \cr
\Delta_{{\kappa}}(K_{ij})&=&\Delta_0(K_{ij})+
\frac{i}{2{{\kappa}}}\left[K_{ij},V_\rho\right]\wedge \Pi_\tau +
\frac{1}{2{{\kappa}}}V_\rho \wedge(\delta_{i\tau}\Pi_j
-\delta_{j\tau}\Pi_i) \;;
 \label{gacoa100}
\end{eqnarray}
\\
the antipodes and counits remain  undeformed ($S_0 (a) = -a$,
$\epsilon
(a) =1$).\\
The deformed coproducts (\ref{gacoa1})-(\ref{gacoa100}) are obtained
by twist procedure \cite{twistpro}, i.e.

\begin{equation}
 \Delta _{\kappa }(a) = \mathcal{F}_{\kappa }\circ
\,\Delta _{0}(a)\,\circ \mathcal{F}_{\kappa
}^{-1}\;\;\;;\;\;\;\Delta _{0}(a) = a \otimes 1 + 1 \otimes
a\;,\label{fs}
\end{equation}
\\
where  the twist factor $\mathcal{F}_{\kappa } \in {\mathcal
U}_{\kappa}(\mathcal{G}) \otimes {\mathcal U}_{\kappa}(\mathcal{G})$
satisfies the classical cocycle condition

\begin{equation}
{\mathcal F}_{{\kappa}12} \cdot(\Delta_{0} \otimes 1) ~{\cal
F}_{\kappa } = {\mathcal F}_{{\kappa }23} \cdot(1\otimes \Delta_{0})
~{\mathcal F}_{{\kappa }}\;, \label{cocyclef}
\end{equation}
\\
and the normalization condition

\begin{equation}
(\epsilon \otimes 1)~{\cal F}_{{\kappa}} = (1 \otimes
\epsilon)~{\cal F}_{{\kappa }} = 1\;\;\;;\;\;\;{\cal F}_{{\kappa
}12} = {\cal F}_{{\kappa }}\otimes 1\;\;,\;\;{\cal F}_{{\kappa }23}
= 1 \otimes {\cal F}_{{\kappa }}\;. \label{normalizationhh}
\end{equation}
\\
It looks as follows

\begin{eqnarray}
{\cal F}_{\kappa} =  \exp \left(\frac{i}{2\kappa}{\Pi_\tau \wedge
V_\rho} \right)\;. \label{factory0}
\end{eqnarray}
\\
Obviously  for deformation parameter $\kappa$ approaching infinity
the above Hopf structure becomes classical.

Let us now turn to  the deformed space-time corresponding to the
Hopf algebra ${\mathcal U}_{\kappa}(\mathcal{G})$. It is defined as
the quantum representation spaces (Hopf module) for quantum Galilei
algebra, with action of the deformed symmetry generators satisfying
suitably deformed Leibnitz rules (see e.g. \cite{bloch}).  The
action of Galilei group ${\mathcal U}_{\kappa}(\mathcal{G})$ on a
Hopf module of functions depending on space-time coordinates
$(t,x_i)$ is given by

\begin{eqnarray}
&&\Pi_{0}\rhd
f(t,\overline{x})=i{\partial_t}f(t,\overline{x})\;\;\;,\;\;\;
\Pi_{i}\rhd f(t,\overline{x})=i{\partial_i}f(t,\overline{x})\;,
\label{a1}\\
&~~&  \cr &&K_{ij}\rhd f(t,\overline{x}) =i\left( x_{i }{\partial_j}
-x_{j }{\partial_i} \right) f(t,\overline{x})\;\;\;,\;\;\; V_i\rhd
f(t,\overline{x}) =it{\partial_i}
\,f(t,\overline{x})\;.~~~~~~~~\label{dsf}
\end{eqnarray}
\\
Moreover, the $\star$-multiplication of arbitrary two functions  is
defined as follows

\begin{equation}
f(t,\overline{x})\star_{{\kappa}} g(t,\overline{x}):=
\omega\circ\left(
 \mathcal{F}_{\kappa}^{-1}\rhd  f(t,\overline{x})\otimes g(t,\overline{x})\right) \;.
\label{star}
\end{equation}
\\
In the above formula $\mathcal{F}_{\cdot}$ denotes  twist factor
(\ref{factory0}) and $\omega\circ\left( a\otimes b\right) = a\cdot
b$.\\
Consequently, we get the following factor

\begin{equation}
\mathcal{F}_{{\kappa}}=  {\rm \exp}
\,\left(-\frac{i}{2{\kappa}}\,\partial_{\tau}\wedge t{\partial_\rho}
\right)\;, \label{swar1}
\end{equation}
\\
and the corresponding nonrelativistic space-time (see
(\ref{wstep1}))

\begin{equation}
[\;t,{ { x}}_{i}\;]_{\star_{{\kappa}}}   = [\;{ { x}}_{k},{{
x}}_{\rho}\;]_{\star_{{\kappa}}} = [\;{ { x}}_{k},{{
x}}_{\tau}\;]_{\star_{{\kappa}}} = 0 \;\;\;,\;\;\;[\;{ {
x}}_{\tau},{ { x}}_{\rho}\;]_{\star_{{\kappa}}} =
\frac{i}{\kappa}t\;,
 \label{ysesstar}
\end{equation}
\\
with indexes $\rho$, $\tau$ and $k$ different and fixed, and $[\,
a,b\,] _{\star_{{\kappa}}}:=a{\star_{{\kappa}}} b -
b{\star_{{\kappa}}}a$.

 Finally, it should be mentioned  that for the deformation
parameter $\kappa$ running to  infinity the above quantum space
becomes commutative.

\section{{{Classical oscillator  model}}}

Let us start with the following Lie-algebraically deformed phase
space corresponding to the quantum space-time (\ref{ysesstar}) (see
\cite{daszwal})\footnote{We use the correspondence relation
$\{\;a,b\;\} = \frac{1}{i}[\;\hat{a},\hat{b}\;]$  $(\hbar = 1)$.}

\begin{equation}
\{\;t,{ {\bar x}}_{i}\;\} = 0 = \{\;{ {\bar x}}_{k},{{\bar
x}}_{\rho(\tau)}\;\}\;\;\;,\;\;\;\{\;{ {\bar x}}_{\tau},{ {\bar
x}}_{\rho}\;\} = \frac{1}{\kappa}t\;, \label{in2}
\end{equation}
\begin{equation}
\{\;{ {\bar x}}_{i},{\bar p}_j\;\} = \delta_{ij}\;\;\;,\;\;\;\{\;{
{\bar p}}_{i},{ {\bar p}}_{j}\;\} = 0\;, \label{in2a}
\end{equation}
\\
where  indices $k$, $\rho$ and $\tau$ are different and fixed, $i,j
= 1,2,3$. One can check that the  relations (\ref{in2}),
(\ref{in2a}) satisfy the Jacobi identity and for deformation
parameter $\kappa$ running to infinity become classical.

We define the Hamiltonian function for isotropic harmonic oscillator
with constant mass $m$ and frequency $\omega$ as follows

\begin{equation}
{\bar{H}}({\bar p},{\bar x})=\frac{1}{2m}\left({\bar p}_{\rho}^2 +
{\bar p}_{\tau}^2 + {\bar p}_{k}^2\right) +
\frac{m\omega^2}{2}\left({\bar x}_{\rho}^2 + {\bar x}_{\tau}^2 +
{\bar x}_{k}^2\right)\;. \label{hamosc}
\end{equation}
\\
Next, in order to analyze the above system we represent the
noncommutative variables $({\bar x}_i, {\bar p}_i)$ on classical
phase space $({ x}_i, { p}_i)$ as  (see e.g. \cite{lumom},
\cite{kijanka}, \cite{giri})

\begin{equation}
{\bar x}_{\rho} = { x}_{\rho} + \frac{t}{2\kappa}
p_\tau\;\;\;,\;\;\;{\bar x}_{\tau} = { x}_{\tau} -\frac{t}{2\kappa}
p_\rho\;\;\;,\;\;\; {\bar x}_{k}= x_k \;\;\;,\;\;\; {\bar p}_{i}=
p_i\;, \label{rep}
\end{equation}
\\
where

\begin{equation}
\{\;x_i,x_j\;\} = 0 =\{\;p_i,p_j\;\}\;\;\;,\;\;\; \{\;x_i,p_j\;\}
=\delta_{ij}\;. \label{classpoisson}
\end{equation}
\\
Then, the  Hamiltonian (\ref{hamosc}) takes the form

\begin{eqnarray}
{{H}}({ p},{ x})=H(t) =\frac{\left({ p}_{\rho}^2 + {
p}_{\tau}^2\right)}{2M(t)} +\frac{M(t)\Omega^2(t)}{2}\left({
x}_{\rho}^2 +
{x}_{\tau}^2 \right) 
+\frac{tm\omega^2L_k}{2\kappa}
+\frac{{ p}_{k}^2}{2m} +\frac{m\omega^2{ x}_{k}^2}{2}\label{hamoscnew}
\end{eqnarray}
\\
with symbol $L_k = x_\rho p_\tau - x_\tau p_\rho$ denoting angular
momentum of
 particle in direction $k$. Besides, 
 the coefficients
 $M(t)$ and $\Omega (t)$ present in the  above formula  denote the
 time-dependent  functions given by\footnote{$\lim\limits_{t\to\infty}M(t)=0$,
$\lim\limits_{t\to\infty}\Omega (t)= \infty$.}, \footnote{See Fig.
2a.}

\begin{equation}
M(t)= \frac{m}{1 + \frac{m^2\omega^2t^2}{4\kappa^2}}
\;\;\;,\;\;\;\Omega (t)= \omega \sqrt{1 +
\frac{m^2\omega^2t^2}{4\kappa^2}}\;, \label{massfre}
\end{equation}
\\
respectively. One can check that 

\begin{equation}
M(t) \Omega^2 (t)= m \omega^2 = {\rm const.}\;. \label{prawo}
\end{equation}

As it was already mentioned in Introduction, the considered  system
(\ref{hamoscnew})
 with neglected
 last three terms as well as with   arbitrary "mass" and "frequency" functions $M(t)$, $\Omega (t)$   has
 been  studied at  classical and quantum levels in
 \cite{slawia}-\cite{slawia10}. Obviously, in our case, due to
 the condition (\ref{prawo}) the coefficient of a second term in
 Hamiltonian function (\ref{hamoscnew}) remains constant, i.e.

\begin{equation}
 \frac{M(t)\Omega^2(t)}{2}\left({ x}_{\rho}^2 +
{x}_{\tau}^2 \right) = \frac{m\omega^2}{2}\left({ x}_{\rho}^2 +
{x}_{\tau}^2 \right)\;.\label{slowianszczyzna}
\end{equation}

\begin{figure}[]
\begin{center}
\includegraphics[width=15.5cm]{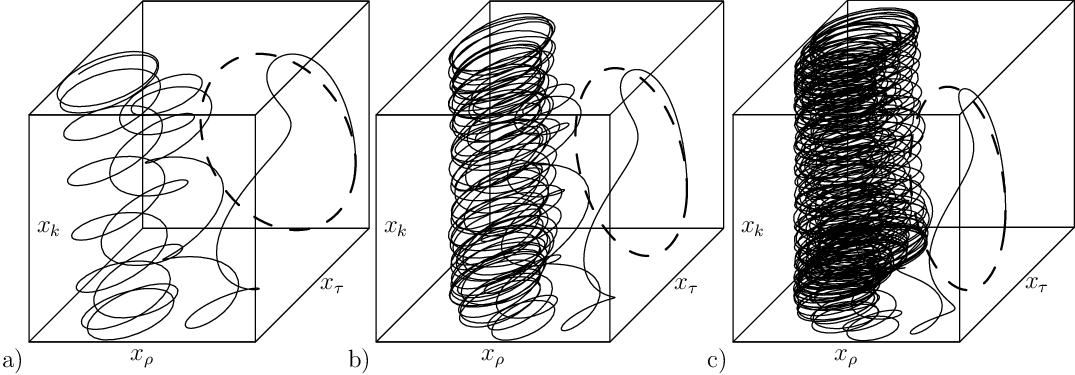}
\end{center}
\caption{The particle trajectory for parameters $m=\omega=\kappa=1$.
The dashed line corresponds to the  case of classical (undeformed)
oscillator and the time parameter runs from 0 to 15, 30 and 50 for
figures a), b) and c) respectively.} \label{fig.0}
\end{figure}

Using the formulas (\ref{classpoisson}), (\ref{hamoscnew}) one gets
the following canonical Hamiltonian equations of motions

\begin{eqnarray}
&&\dot{x}_{\rho} = \frac{p_\rho}{M(t)} - \frac{tm\omega^2
}{2\kappa}x_\tau \;\;\;,\;\;\; \dot{p}_{\rho} = -m\omega^2 x_\rho -
\frac{tm\omega^2 }{2\kappa}p_\tau
\;,\label{ham1}\\
 &~~&~\cr
&&\dot{x}_{\tau} = \frac{p_\tau}{M(t)} + \frac{tm\omega^2
}{2\kappa}x_\rho\;\;\;,\;\;\; \dot{p}_{\tau} = -m\omega^2x_\tau +
\frac{tm\omega^2}{2\kappa}p_\rho
\;,\label{ham2}\\
&~~&~\cr 
&&~~~~~~~~~~\dot{x}_{k} = \frac{p_k}{m}\;\;\;,\;\;\;\dot{p}_{k} =
-m\omega^2 x_k\;,\label{ham3}
\end{eqnarray}
\\
which when combined yield the equations

\begin{equation}
\left\{\begin{array}{rcl} \ddot{x}_\rho  &=&
\dfrac{m\omega^2}{2\kappa} t\left(-2\dot{x}_\tau+
\dfrac{M(t)}{\kappa}\dot{x}_\rho\right)+\dfrac{m\omega^2}{2\kappa}
\left(\dfrac{m\omega^2}{2\kappa^2}M(t)t^2-1\right)x_\tau-\omega^2 x_\rho\\
 &~~&~\cr
 \ddot{x}_\tau  &=&
\dfrac{m\omega^2}{2\kappa}t\left(2\dot{x}_\rho+
\dfrac{M(t)}{\kappa}\dot{x}_\tau\right)+\dfrac{m\omega^2}{2\kappa}
\left(1-\dfrac{m\omega^2}{2\kappa^2}M(t)t^2\right)x_\rho-\omega^2 x_\tau\\
 &~~&~\cr
 \ddot{x}_k  &=&-\omega^2 x_k
 \;.\end{array}\right.\label{dddmixednewton1}
\end{equation}
\\
The solution of eq. (\ref{dddmixednewton1}) has been studied
numerically
and  the corresponding trajectories are illustrated on Fig.
\ref{fig.0}. If the parameter ${{\kappa}}$ runs to infinity the
equation (\ref{dddmixednewton1}) becomes undeformed and describes
the periodic motion of classical harmonic oscillator
\cite{ksiazkamech}.

\section{{{Quantum oscillator model}}}

The main aim of this Section is to study the spectrum of the
following quantum-mechanical  counterpart of  Hamiltonian  (\ref{hamoscnew})

\begin{eqnarray}
{\hat{H}}(t) =\frac{\left({ \hat p}_{\rho}^2 + {\hat
p}_{\tau}^2\right)}{2M(t)} +\frac{m\omega^2}{2}\left({\hat
x}_{\rho}^2 + {\hat x}_{\tau}^2\right) +\frac{tm\omega^2{\hat
L}_k}{2\kappa} +\frac{{\hat p}_{k}^2}{2m} +\frac{m\omega^2{\hat
x}_{k}^2}{2} \;, \label{hamquant}
\end{eqnarray}
\\
with  $\hat{x}_i$ and $\hat{p}_i$ denoting the classical  position
and momentum operators such that

\begin{equation}
[\;\hat{x}_i,\hat{x}_j\;] =0 =[\;\hat{p}_i,\hat{p}_j\;]
\;\;\;\;,\;\;\;\; [\;\hat{x}_i,\hat{p}_j\;] =i\delta_{ij}
\;.\label{ccr}
\end{equation}
\\
In accordance with the scheme proposed in  \cite{kijanka} (see also
\cite{lumom}) we introduce a set of time-dependent creation
$(a^{\dag}_{A}(t))$ and annihilation
$(a_{A}(t))$ operators 

\begin{eqnarray}
\hat{a}_{\pm}(t) &=& \frac{1}{2}\left[\frac{(\hat{p}_\rho \pm
i\hat{p}_\tau)}{\sqrt{M(t)\Omega (t)}} -i\sqrt{M(t)\Omega
(t)}(\hat{x}_\rho \pm
i\hat{x}_\tau)\right]\;,\label{oscy1}\\
&~~&~\cr \hat{a}_{k} &=&
\frac{1}{\sqrt{2}}\left(\frac{\hat{p}_k}{\sqrt{m\omega }}
-i\sqrt{m\omega }\hat{x}_k\right)\;,\label{oscy2}
\end{eqnarray}
\\
 satisfying the standard commutation relations

\begin{eqnarray}
[\;\hat{a}_{A},\hat{a}_{B}\;] =
0\;\;,\;\;[\;\hat{a}^{\dag}_{A},\hat{a}^{\dag}_{B}\;]
=0\;\;,\;\;[\;\hat{a}_{A},\hat{a}^{\dag}_{B}\;] =
\delta_{AB}\;\;\;;\;\;A,B = \pm, k \;.\label{ccr1}
\end{eqnarray}
\\
It should be noted,  that for the parameter $\kappa$ running to
infinity the creation/annihilation operators (\ref{oscy1}) take the
classical form,  i.e. they become time-independent. Further,  one
can easily check that in  terms of the  operators (\ref{oscy1}) and
(\ref{oscy2}) the Hamiltonian function (\ref{hamquant}) looks as
follows

\begin{equation}
{\hat{H}}(t)=\Omega_{+}(t) \left({\hat N}_+(t) + \frac{1}{2}\right)
+ \Omega_{-}(t) \left({\hat N}_-(t) + \frac{1}{2}\right) + \omega
\left({\hat N}_k + \frac{1}{2}\right) \;, \label{hamquantosc}
\end{equation}
\\
with  coefficient  $\Omega_{\pm}(t)$  given by\footnote{See Fig.
2b.}

\begin{equation}
\Omega_{\pm}(t)=\Omega(t)\mp \frac{t{m\omega^2
}}{2\kappa}\;.\label{ompm}
\end{equation}
\\
Besides, the following objects

\begin{equation}
{\hat N}_{\pm}(t)={\hat a}^{\dag}_{\pm}(t){\hat
a}_{\pm}(t)\;\;\;,\;\;\; {\hat N}_{k}={\hat a}^{\dag}_{k}{\hat
a}_{k}\;,\label{nn}
\end{equation}
\\
play a role of particle number operators in $\pm$ and $k$ direction
respectively.

The eigenvectors of Hamiltonian (\ref{hamquantosc}) are generated by
creation operators $a^{\dag}_{\pm}(t)$, $a^{\dag}_{k}$ acting on
vacuum state $|0>$

\begin{eqnarray}
|n_+,n_-,n_k>_{t} =
\frac{1}{\sqrt{n_+!}}\frac{1}{\sqrt{n_-!}}\frac{1}{\sqrt{n_k!}}\left({\hat
a}^{\dag}_{+}(t)\right)^{n_+} \left({\hat
a}^{\dag}_{-}(t)\right)^{n_-}\left({\hat
a}^{\dag}_{k}\right)^{n_k}|0>\;.\label{state}
\end{eqnarray}
\\
Then, the corresponding eigenvalues take the form

\begin{equation}
E_{n_+,n_-,n_k}(t) = \Omega_{+}(t) \left(n_+ + \frac{1}{2}\right) +
\Omega_{-}(t) \left(n_- + \frac{1}{2}\right) + \omega \left(n_k +
\frac{1}{2}\right)\;. \label{eigenvalues}
\end{equation}
\\
One can also see that the difference between two neighboring levels
of spectrum (\ref{eigenvalues}) is 
given by the formula

\begin{equation}
\Delta E(t) := E_{n_++1,n_-+1,n_k+1}(t) - E_{n_+,n_-,n_k}(t)
=2\Omega (t)+ \omega\;, \label{diff}
\end{equation}
\\
with function $\Delta E(t)$  growing  in time.  
Obviously, for deformation parameter $\kappa$ running to infinity
the spectrum (\ref{eigenvalues}) as well as its  difference
(\ref{diff}) become classical. As an  illustration of relation
(\ref{eigenvalues}) the time evolution of
$E_{0,0,0}(t)$ and $E_{1,1,1}(t)$  eigenvalues 
has been presented on Fig. 2c.

Finally, it should be noted that for fixed parameter $t$ the above
system can be identified with canonically deformed oscillator model
proposed in \cite{kijanka}, i.e. the spectrum (\ref{eigenvalues})
for time parameter $t = \kappa \theta_{\rho \tau}$ (see
(\ref{wielkaslawia})) becomes   the same as  one derived in
\cite{kijanka}.

\begin{figure}[]
\begin{center}
\includegraphics[width=14.5cm]{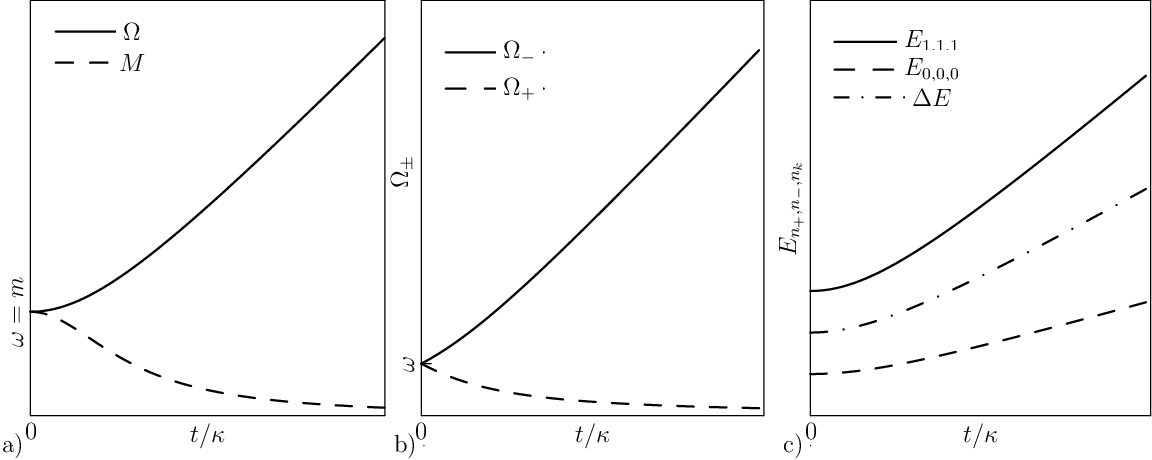}
\end{center}
\caption{The time evolution of functions $\Omega(t)$, $M(t)$,
$\Omega_{\pm}(t)$, $E_{0,0,0}(t)$, $E_{1,1,1}(t)$ and difference
$\Delta E(t) =E_{1,1,1}(t) - E_{0,0,0}(t)$ with fixed   parameter
$\kappa$ and  $m=\omega=1$.} \label{fig.1}
\end{figure}

\section{{{Beyond the quantum group}}}

Let us consider the following generalized  phase space

\begin{equation}
\{\;t,{ {\bar x}}_{i}\;\} = 0 = \{\;{ {\bar x}}_{k},{{\bar
x}}_{\rho(\tau)}\;\}\;\;\;,\;\;\;\{\;{ {\bar x}}_{\tau},{ {\bar
x}}_{\rho}\;\}  = f_{\kappa}(t)\;, \label{beyond}
\end{equation}
\begin{equation}
\{\;{ {\bar x}}_{i},{\bar p}_j\;\} = \delta_{ij}\;\;\;,\;\;\;\{\;{
{\bar p}}_{i},{ {\bar p}}_{j}\;\} = 0\;, \label{genin2a}
\end{equation}
\\
where $f_{\kappa}(t)$ denotes arbitrary time-dependent function
approaching to zero  for  parameter $\kappa$ running to infinity.
Obviously, the above relations satisfy the Jacobi identity and for
parameter $\kappa$ approaching to infinity become classical.
Besides, as it was already mentioned in Introduction, the
generalized noncommutativity (\ref{beyond}) can not be  realized as
a translation sector in the Hopf-algebraic framework of relativistic
and nonrelativistic symmetries. Nevertheless, due to the link (for
particular choices of function $f_{\kappa}(t)$) with oscillator
models \cite{slawia10a}, \cite{slawia10}, the study of such a system
appears quite interesting and   shall be discussed in present
Section.

The relations (\ref{beyond}) and (\ref{genin2a}) can be represented
in terms of  classical phase space variables as follows

\begin{equation}
{\bar x}_{\rho} = { x}_{\rho} + \frac{f_{\kappa}(t)}{2}
p_\tau\;\;\;,\;\;\;{\bar x}_{\tau} = { x}_{\tau}
-\frac{f_{\kappa}(t)}{2} p_\rho\;\;\;,\;\;\; {\bar x}_{k}= x_k
\;\;\;,\;\;\; {\bar p}_{i}= p_i\;. \label{genrep}
\end{equation}
\\
Then,  the corresponding Hamiltonian function   takes the form

\begin{eqnarray}
H_f(t) =\frac{\left({ p}_{\rho}^2 + { p}_{\tau}^2\right)}{2M_f(t)}
+\frac{m\omega^2}{2}\left({ x}_{\rho}^2 +
{x}_{\tau}^2\right)+\frac{f_{\kappa}(t)m\omega^2L_k}{2} +\frac{{
p}_{k}^2}{2m} +\frac{m\omega^2{ x}_{k}^2}{2}\;, \label{genhamoscnew}
\end{eqnarray}
\\
with generalized coefficient $M_f(t)$ 
  given by

\begin{equation}
M_f(t)= \frac{m}{1 + \frac{m^2\omega^2}{4}f_{\kappa}^2(t) } \;.
\label{genmassfre}
\end{equation}
\\
By direct calculation one can also find the corresponding  equation
of motion

\begin{equation}
\left\{\begin{array}{rcl} \ddot{x}_\rho &=& \dfrac{m\omega^2
f_\kappa(t)}{2}\left( \dot{f}_\kappa(t)M_f(t) \dot{x}_\rho
-2\dot{x}_\tau\right)+\\
&~~&~\cr &&~~~~~~~~~~~~~~~~~~+ \dfrac{m\omega^2\dot{f}_\kappa(t)}{2}
\left(\dfrac{m\omega^2 M_f(t)}{2}f^2_\kappa(t)-1\right)x_\tau-\omega^2 x_\rho\\
 &~~&~\cr
 \ddot{x}_\tau  &=&
\dfrac{m\omega^2 f_\kappa(t)}{2}\left( \dot{f}_\kappa(t)M_f(t)
\dot{x}_\tau +2\dot{x}_\rho\right)+\\
&~~&~\cr &&~~~~~~~~~~~~~~~~~~+ \dfrac{m\omega^2\dot{f}_\kappa(t)}{2}
\left(1-\dfrac{m\omega^2 M_f(t)}{2}f^2_\kappa(t)\right)x_\rho-\omega^2 x_\tau\\
 &~~&~\cr
 \ddot{x}_k  &=&-\omega^2 x_k
 \;,\end{array}\right.\label{gendddmixednewton1}
\end{equation}
\\
which in the case $f_\kappa(t)= \frac{t}{\kappa}$ leads to the
 system (\ref{in2}), (\ref{in2a}) related with quantum group
 \cite{daszgali}. Besides, for the choice $f_\kappa(t)= \theta^{\rho\tau}$ we
get the equations of motion for canonical deformation of Galilei
algebra proposed in \cite{romeroads}.

\begin{figure}[]
\begin{center}
\includegraphics[width=15.5cm]{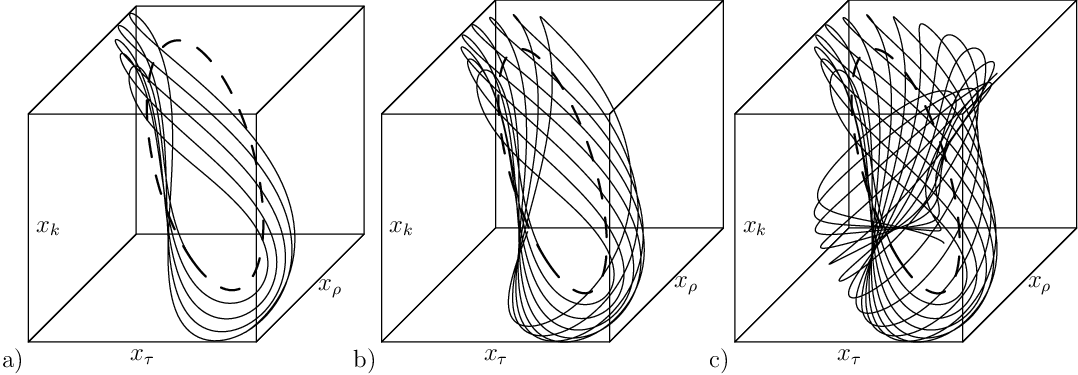}
\end{center}
\caption{The particle trajectory for function  ${f}_\kappa(t) = \sin
\left(\frac{t}{\kappa}\right)$ and parameters  $m=\omega=\kappa=1$.
The dashed line corresponds to the  case of classical (undeformed)
oscillator and the time parameter runs from 0 to 30, 50 and 100 for
figures a), b) and c) respectively.} \label{fig.01}
\end{figure}

\begin{figure}[]
\begin{center}
\includegraphics[width=15.5cm]{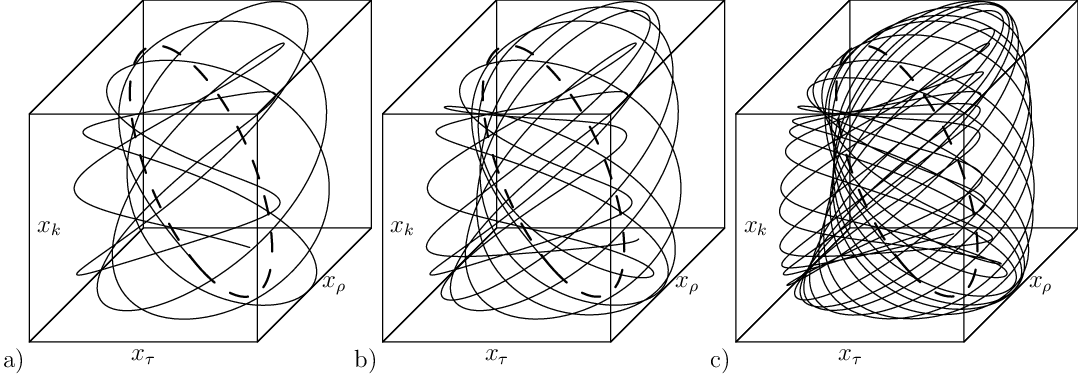}
\end{center}
\caption{The particle trajectory for function  ${f}_\kappa(t) =
\left( {\rm e}^{-\frac{t}{\kappa}}-1\right)$ and parameters
$m=\omega=\kappa=1$. The dashed line corresponds to the case of
classical (undeformed) oscillator and the time parameter runs from 0
to 30, 50 and 100 for figures a), b) and c) respectively.}
 \label{fig.02}
\end{figure}

The above equations have been investigated numerically for two
particular choices of function $f_{\kappa}(t)$ with fixed parameter
$\kappa$

\begin{equation}
{f}_\kappa(t) = \sin \left(\frac{t}{\kappa}\right)\;\;({\rm
a})\;\;\;\;{\rm and}\;\;\;\; {f}_\kappa(t) = \left( {\rm
e}^{-\frac{t}{\kappa}}-1\right)\;\;({\rm b})\;.\label{choice}
\end{equation}
\\
First of them corresponds to the periodic time evolution of
generalized coefficient (\ref{genmassfre}) and can be compare with
the oscillator system  studied in \cite{slawia10a},
\cite{slawia10}\footnote{More preciously, there was considered in
\cite{slawia10a}, \cite{slawia10} the oscillator model described by
the Hamiltonian function (\ref{genhamoscnew}) with neglected $L_k$-,
$x_k$- and $p_k$-terms as well as with time-periodic coefficient
$M_f(t)$ $(m
=M_f)$ and constant frequency $\omega$.}. The second choice  
introduces, as we shall see below,  the finite  for large times
energy spectrum of a proper quantum oscillator model.
Of course,  for $\kappa \to \infty$ the both functions (35a) and
(35b) approach to zero, and the phase space (\ref{beyond}),
(\ref{genin2a}) becomes classical. We add that the corresponding
trajectories are illustrated on Fig. \ref{fig.01} and \ref{fig.02}.


Let us now turn to the energy spectrum of quantum oscillator model
defined on generalized
phase space (\ref{beyond}), (\ref{genin2a}). 
If we introduce the set of the following generalized
creation/annihilation operator

\begin{eqnarray}
\hat{a}_{f\pm}(t) &=& \frac{1}{2}\left[\frac{(\hat{p}_\rho \pm
i\hat{p}_\tau)}{\sqrt{M_f(t)\Omega_f (t)}} -i\sqrt{M_f(t)\Omega_f
(t)}(\hat{x}_\rho \pm
i\hat{x}_\tau)\right]\;,\label{genoscy1}\\
&~~&~\cr \hat{a}_{k} &=&
\frac{1}{\sqrt{2}}\left(\frac{\hat{p}_k}{\sqrt{m\omega }}
-i\sqrt{m\omega }\hat{x}_k\right)\;,\label{genoscy2}
\end{eqnarray}
\\
then, in  analogy to  Section 2, one  gets 

\begin{equation}
E_{f\,n_+,n_-,n_k}(t) = \Omega_{f+}(t) \left(n_+ +
\frac{1}{2}\right) + \Omega_{f-}(t) \left(n_- + \frac{1}{2}\right) +
\omega \left(n_k + \frac{1}{2}\right)\;, \label{geneigenvalues}
\end{equation}
\\
with time-dependent  coefficients

\begin{equation}
\Omega_{f\pm}(t)=\Omega_f(t)\mp
\frac{f_{\kappa}(t){m\omega^2}}{2}\;\;\;\;\;{\rm
and}\;\;\;\;\;\Omega_f (t)= \omega \sqrt{1 +
\frac{m^2\omega^2}{4}f_{\kappa}^2 (t)}\;.\label{genompm}
\end{equation}
\\
Additionally, one can  observe that

\begin{equation}
\Delta E_f(t) := E_{f\,n_++1,n_-+1,n_k+1}(t) - E_{f\,n_+,n_-,n_k}(t)
=2\Omega_f (t) + \omega\;. \label{gendiff}
\end{equation}

\begin{figure}[]
\begin{center}
\includegraphics[width=15.5cm]{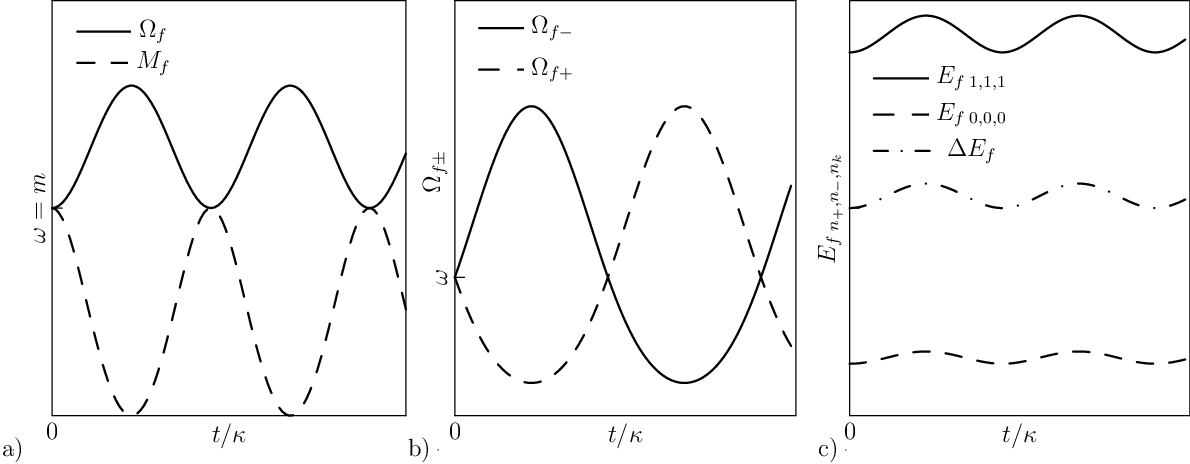}
\end{center}
\caption{The time evolution of functions $\Omega_f(t)$, $M_f(t)$,
$\Omega_{f\pm}(t)$, $E_{f\,0,0,0}(t)$, $E_{f\,1,1,1}(t)$ and
difference $\Delta E_f(t) = E_{f\,1,1,1}(t)- E_{f\,0,0,0}(t)$ for
function $f_\kappa(t)=\sin(t/\kappa)$ with fixed  parameter $\kappa$
and  $m=\omega=1$.}
 \label{fig.2}
\end{figure}

\begin{figure}[]
\begin{center}
\includegraphics[width=15.5cm]{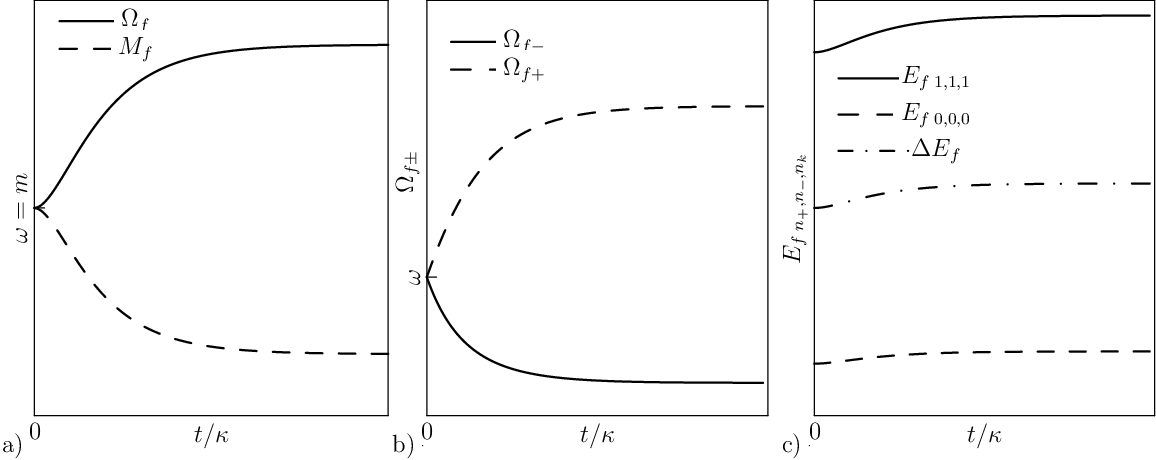}
\end{center}
\caption{The time evolution of functions $\Omega_f(t)$, $M_f(t)$,
$\Omega_{f\pm}(t)$, $E_{f\,0,0,0}(t)$, $E_{f\,1,1,1}(t)$ and
difference $\Delta E_f(t)= E_{f\,1,1,1}(t)- E_{f\,0,0,0}(t)$ for
function $f_\kappa(t)=(\mathrm{e}^{-t/\kappa}-1)$ with fixed
parameter $\kappa$ and  $m=\omega=1$.} \label{fig.3}
\end{figure}

Finally, let us note that in the case of function (35a) with fixed
parameter $\kappa$,
the formula (\ref{genompm}) takes the  form

\begin{equation}
\Omega_f (t)= \omega \sqrt{1 + \frac{m^2\omega^2}{4}\sin^2
\left(\frac{t}{\kappa}\right)}\;, \label{example1}
\end{equation}
\\
and   difference (\ref{gendiff}) becomes
 periodic in time  (see Fig. \ref{fig.2}).
 For the second, "exponential" choice of $f_{\kappa}(t)$, the function

\begin{equation}
\Omega_f (t)= \omega \sqrt{1 + \frac{m^2\omega^2}{4}\left( {\rm
e}^{-\frac{t}{\kappa}}-1\right)^2}\;, \label{example2}
\end{equation}
\\
 is smaller than $\Omega_f
= \omega \sqrt{1 + \frac{m^2\omega^2}{4}}$ for large times, 
and then, the difference (\ref{gendiff}) becomes
 finite (see Fig. \ref{fig.3}). The time evolution of $E_{0,0,0}(t)$ and
$E_{1,1,1}(t)$ eigenvalues has been presented for two considered
cases on Fig. \ref{fig.2} and \ref{fig.3}  respectively.


\section{{{Final Remarks}}}

In this article we investigate the classical and quantum oscillator
model defined on noncommutative space-time (\ref{wstep1}) and its
generalized version (\ref{wstep2}). The corresponding equations of
motion are provided and the time-dependent spectra of both quantum
models are analyzed and illustrated.

As we already mentioned, the presented investigations describe the
link between noncommutative space-time geometry and oscillator
models with time-dependent mass and frequency
\cite{slawia}-\cite{slawia10}.  We mentioned that better
understanding of such a connection based on  more detailed studies
 is postponed for future investigations.

It should be noted that the present project can be extended in
various ways. First of all, one should consider  more complicated
system like a particle in a central field potential defined on
quantum spaces (\ref{wstep1}) and (\ref{wstep2}). Secondly, one can
investigate the oscillator
model on other 
deformed space-times such as a fuzzy  or twisted 
quantum spaces \cite{fuzzy} and \cite{daszgali}. Finally, one should
ask about oscillator system defined on phase space with position
noncommutativity (\ref{in2}), (\ref{beyond}) supplemented by
suitable deformation of the momentum sector. Such a model has been
considered recently in \cite{giri} in a context of noncommutative
space-time (\ref{wielkaslawia}) with additionally deformed Poisson
brackets for momentum coordinates $p_\mu$. The studies in these
directions  are in progress.

\section*{Acknowledgments}
The authors would like to thank J. Lukierski, Z. Haba, J.
Jedrzejewski, R. Olkiewicz, Z. Petru, Z. Popowicz,  L. Turko and M.
Woronowicz
for valuable discussions.\\
This paper has been financially supported by Ministry of Science and
Higher Education grant NN202318534.

\end{document}